\documentclass[preprint,12pt]{elsarticle}

\usepackage{amssymb}
\usepackage{slashbox}
\usepackage{mathrsfs}
\usepackage{mathrsfs}
\usepackage{mathrsfs}
\usepackage{amsfonts}
\usepackage{amsfonts}
\usepackage{amsfonts}
\usepackage{mathrsfs}
\usepackage{amsmath}
\usepackage{tabularx}
\usepackage{amsmath}

\usepackage{graphicx}
\usepackage{multirow}
\usepackage{caption2}
\usepackage{float}
\usepackage{booktabs}
\usepackage{algorithm}
\usepackage{algorithmic}
\usepackage{subfigure}
\usepackage{color}
\usepackage{makecell}
\numberwithin{equation}{section}

\begin{document}

\begin{frontmatter}



\title{Re-scale AdaBoost for Attack Detection in Collaborative Filtering Recommender Systems \tnoteref{t1}} \tnotetext[t1]{
The research was
supported by the National 973 Programming (2013CB329404) and the National Natural Science Foundation (Grant No.
11131006, 11401462 and 61221063).
}

\author{Zhihai~Yang$^1$  }

\author{Lin Xu$^{2}$\corref{*}}\cortext[*]{Corresponding author: xulinshadow@gmail.com}

\author{Zhongmin~Cai$^1$}

%\author{Zongben Xu$^1$}

\address{1. Ministry of Education Key Lab for Intelligent Networks and Network Security,
Xi'an Jiaotong University, Xi'an, 710049, China

2. Institute for Information and System Sciences, School of Computer Science and Technology,
Xi'an Jiaotong University, Xi'an, 710049, China}

\begin{abstract}
Collaborative filtering recommender systems (CFRSs) are the key components of successful e-commerce systems.
Actually, CFRSs are highly vulnerable to attacks since its openness.
However, since attack size is far smaller than that of genuine users, conventional supervised learning based detection
methods could be too ``dull'' to handle such imbalanced classification.
In this paper, we improve detection performance from following two aspects.
First, we extract well-designed features from user profiles based on the
statistical properties of the diverse attack models, making hard classification task becomes easier to perform.
Then, refer to the general idea of  re-scale Boosting (RBoosting) and AdaBoost, we apply a variant of AdaBoost, called the re-scale AdaBoost (RAdaBoost) as our detection method based on extracted features. RAdaBoost is comparable to the optimal Boosting-type algorithm and can effectively improve the performance in some hard scenarios.
Finally, a series of experiments on the MovieLens-100K data set are conducted to demonstrate the outperformance of RAdaBoost comparing with some
classical techniques such as SVM, kNN and AdaBoost.
\end{abstract}
\end{frontmatter}
% Introduction
%%%%%%%%%%%%%%%%%%%%%%%%%%%%%%%%%%%%%%%%%%%%%%%%%%%%
\section{Introduction}
Personalization  recommender systems (RSs) utilize a variety of recommendation methods to suggest products that users may like, such as movies, music, news, books and other products.
Collaborative filtering recommender systems (CFRSs) have been proved to be one of the most successful RSs  used  by  many  e-commerce  companies  such as Amazon, Ringo, eBay, GroupLens etc. \cite{Bryan2008}, \cite{Burke2006}, \cite{Chung2013}, \cite{Mehta2007}.
In practice, CFRSs are prone to manipulation from attackers since its openness. Typically, attackers carefully inject chosen attack profiles into CFRSs in order to bias the recommendation results to their benefits, which is termed ``shilling'' or ``profile injection''
attacks.
It decreases the trustworthiness of recommendation and leads to a negative impact on the CFRSs.
Thus, constructing an effective method to defend the attackers and remove them from the CFRSs is crucial.

Supervised learning based detection method for ``shilling'' or ``profile injection'' in CFRSs is  an important research direction, which regards the detection attributions as the classification features and distinguishes  attack profiles from  genuine profiles by constructed features.
Actually, the attack detection problem can be formulated as an imbalanced classification. The number of attackers is far smaller than genuine users in CFRSs, especially when attack size \footnote{The ratio between the number of attackers and genuine users.} is small. However, the traditional supervised  learning (i.e., SVM and kNN) based  attack  detection  methods often inevitably have individual weaknesses for handling this kind of issues and fail to effectively capture the concerned attackers.

In the current paper, we aim to improve  detection performance from two aspects.
Firstly, we consider the overall statistical signature of attack profiles
would differ significantly from that of genuine profiles. The
difference comes from  two sources: the distribution of ratings (or
items among the filler items or selected items) and the ratings of the target items.
Based on the statistical properties of the diverse attack models, as many extracted features as possible are designed and used to transform their ``inputs'', distorting the space so that the task (i.e., classification or clustering) becomes easier to perform.
Specifically, we extract as many as 18 features from user profiles (consists of attack profiles and genuine profiles) to construct a sophisticated features representation for each user to make it much more easily classified.
Secondly, refer to the general idea of  re-scale Boosting (RBoosting) \cite{Lin2015}, \cite{Xu2015b} and AdaBoost \cite{Freund1995a, Freund1996}, we apply a variant of Boosting algorithm, called the re-scale AdaBoost (RAdaBoost) as our detection method based on extracted features.
RBoosting is theoretically and experimentally  proved to be better than the classical Boosting algorithm \cite{Lin2015}. Furthermore, the theoretical near optimality of the numerical convergence of RBoosting among all the variants of the Boosting-type algorithms was also specified. This means that if the parameter is appropriately selected, RBoosting is comparable to the optimal Boosting-type algorithm.
And AdaBoost  \cite{Freund1995a, Freund1996} is one of the most popular ensemble techniques paradigm and has been shown to be very effective in practice in some hard scenarios \cite{Harrington2012}. Typically, AdaBoost employs re-weighted loss function for gradually increasing emphasis (or weights) on misclassifications (i.e., concerned attackers) and can distinctly improve the predictive performance on a difficult data set.
Thus, with the help of the re-scale operator,
RAdaBoost can be used in conjunction with many other types of learning algorithms (or weak learners) to improve the performance in ``shilling'' attacks detection.
Finally, a series of experiments on the MovieLens-100K dataset are conducted to demonstrate the outperformance (i.e., classification error, detection rate and false alarm rate) of RAdaBoost comparing with conventional classification techniques such as SVM, kNN and the original non-rescale AdaBoost version.  The experimental results show that RAdaBoost can effectively improve the performance.
%In addition, we also detect the recent published attack models such as power user attack (PUA) \cite{Seminario2014a, Wilson2013, Seminario2013, Wilson2014} and power item attack (PIA) \cite{Seminario2014b} to validate the effectiveness of our proposed approach.

The rest of paper is organized as follows. In Section 2, we
give a brief introduction to the related work. In Section 3, we give a brief introduction of attack profiles and attack models.
In Section 4, our approach are described in details.  In Section 5, experimental results are reported and analyzed.
In the last section, we conclude the paper with a brief summary and prospect the directions of future works.

\section{Related work}

Existing work in this area have focused on detecting and preventing the ``shilling'' attacks (or ``profile injection'' attacks).
Burke et al. \cite{Burke2006} proposed and studied several attributions derived from user profiles for their utility in attack detection. They employed the kNN classifier as their detection approach. But it is unsuccessful when detecting attacks with small filler size \footnote{The ratio between the number of items rated by user $u$  and the number of entire items in the recommender system.}. Then, Williams et al. \cite{Williams2007a}, \cite{Williams2007b} tried to extract features from user profiles and utilized them to detect shilling attacks.
They also suffered from low detection accuracy and many genuine profiles are misclassified as attack profiles. After that, He et al. \cite{He2010} introduced the rough set theory into  shilling attacks detection  by means of taking features of user profiles as the condition attributes of the decision table. However, their method  faced with the low overall classification rate in some cases, especially for bandwagon attack. Afterwards, Wu et al. \cite{Wu2012} proposed a hybrid detection method to detect shilling attacks, which combined the naive Bayesian classifiers and augmented expectation maximization based on several selected metrics. Regretfully, their technique also suffered from low F-measure \cite{David2011} when the filler size is small.
Zhang et al. \cite{Zhang2012} introduced the idea of ensemble learning for improving predictive capability in the attack detection problem.
They constructed the  base-classifiers (or weaker learner) with the Support Vector Machine (SVM) approach and then integrated them to generate a high predictive ability learner for detection. Their proposed method exhibited  better  performance than some benchmarked methods. Nevertheless, it still suffered from low precision  especially when the attack size  is small. In addition, the same authors Zhang et al. \cite{Zhang2014} also proposed an online method, HHT-SVM, to detect profile injection attacks by combining Hilbert-Huang transform (HHT) and support vector machine (SVM). They created rating series for each user profile based on the novelty and popularity of items in order to provide basic data for feature extraction. The precision of their method shown better than the benchmarked methods, but the precision significantly decreased with the filler size increased.

Generally speaking, previous studies showed that the detection results of ``shilling" attacks is dissatisfactory and leave much to be desired, especially when the filler size or attack size is small.
In the current work, we intend to improve the detection performance from two aspects. First, we introduce more well-designed features to depict the distinction between attack profiles and genuine profiles  to make hard classification task (i.e., with small filler size and attack size) becomes easier to perform.
Secondly, in view of conventional classification techniques could be inadequate to handle such imbalanced classification, particularly when the attack size is small,
we applied a variant of AdaBoost algorithm, called the re-scale AdaBoost (RAdaBoost) as our detection method. RAdaBoost gradually increases emphasis on the concerned attacks and can distinctly improve the performance for the imbalanced classification task.

%Gradient boosting machine (GBM) is a strong learning system which combines many parsimonious models to produce a  model with  prominent predictive performance. The underlying intuition is that combines many rough rules of thumb can  yield a good composite learner.
%From the statistical viewpoint,  GBM can be viewed as a paradigm of ``functional gradient decent'' based on any fitting criterion \cite{Friendman2001}. It connects various boosting algorithms to optimization problems with specific loss functions (i.e. $L_2$-Boosting \cite{Buhlmann2003,Freund1996} can be interpreted as an stepwise additive learning scheme that
%concerns the problem of minimizing the $L_2$ risk). Boosting is resistant to overfitting \cite{Friendman2000} and thus, has
%triggered enormous research activities in the past twenty years \cite{Buhlmann2007,Duffy2002,Freund1995,Friendman2001,Schapire1990, Lin2015, Xu2015}.

%The rest of paper is organized as  follows. In Section 2, we
%give a brief introduction to  the $L_2$-Boosting, $L_2$-RBoosting and $L_2$-DDRBoosting.
%In Section 3, we study the related theoretical behaviors of
%$L_2$-RBoosting. In Section 4, a series of simulations and
%real data experiments are employed to illustrate our theoretical assertions. In
%Section 5, we provide the proof of the main results. In the last
%section, we draw a simple conclusion.

\section{Attack profiles and attack models}
The attackers have different attack intents to bias the recommendation results for their benefits. In the literature, ``shilling'' attacks are classified into two ways: nuke attack and push attack \cite{Burke2006}, \cite{Gunes2012}, \cite{Williams2007a}. In nuke attacks, attackers demote the target items by rating the lowest score, whereas in push attacks, attackers promote the target items by rating the highest score.
In order to effectively ``nuke'' or ``push'' a target item, the attacker should  clearly know the form of the attack profiles.
The general form of attack profiles is shown in Table 1. The details of the four sets of items are described as follows:

$I_T$: A set of target items with singleton or multiple items, called single-target attack or multi-target attack. The rating is $\gamma(i_j^T)$, generally rated the maximum or minimum value in the entire profiles.

$I_S$: The set of selected items with specified rating by the function $\sigma(i_k^S)$ \cite{Zhang2013};

$I_F$: A set of filler items, received randomly items with random assigned ratings $\rho(i_l^F)$;

$I_N$: A set of items with no ratings;

In the present work, we utilize 14 attack models to generate attack profiles. The involved attack profiles and corresponding explanations are listed in Table 2.
The details of these attack models are described as follows:

1) Random attack: $I_S=\phi$ and $\rho(i) \sim N( \overline{r}, \overline{\sigma}^2)$ \cite{Zhang2013}.

2) Average attack: $I_S=\phi$ and $\rho(i) \sim N( \overline{r_i}, \overline{\sigma_i}^2)$ \cite{Zhang2013}.

\newcommand{\tabincell}[2]{
\begin{tabular}{@{}#1@{}}#2\end{tabular}
}
\begin{table}[H]
\caption{General form of attack profiles.}
\centering
\renewcommand\arraystretch{1.5}
\begin{tabular}{|c|c|c|c|c|c|c|c|c|c|c|c|}
\hline
\multicolumn{3}{|c|}{$I_T$} & \multicolumn{3}{c|}{$I_S$} & \multicolumn{3}{c|}{$I_F$} & \multicolumn{3}{c|}{$I_N$} \\
\hline
$i_1^T$ & ... & $i_j^T$ & $i_1^S$ & ... & $i_k^S$ & $i_1^F$ & ... & $i_l^F$ & $i_1^N$ & ... & $i_v^N$ \\
\hline
$\gamma(i_1^T)$ & ... & $\gamma(i_j^T)$ & $\sigma(i_1^S)$ & ... & $\sigma(i_k^S)$ & $\rho(i_1^F)$ & ... & $\rho(i_l^F)$ & $null$ & ... & $null$ \\
\hline
\end{tabular}
\end{table}

\begin{table}[H]
\renewcommand{\arraystretch}{1.2}
%\addtolength{\tabcolsep}{-6pt}
%\renewcommand\arraystretch{1.5}
\caption{Attack models summary.}
\begin{center}
\scalebox{0.56}[0.6]{
\begin{tabular}{|c|c|c|c|c|c|c|}
\hline
\multicolumn{1}{|c|}{\multirow{2}{1cm}{Attack Models}}
& \multicolumn{2}{c|}{$I_S$}
& \multicolumn{2}{c|}{$I_F$}
& \multicolumn{1}{c|}{\multirow{2}{*}{$I_N$}}
& \multicolumn{1}{c|}{\multirow{2}{*}{\tabincell{c}{$I_T$ \\ push/nuke}}}
\\
\cline{2-5}
\multicolumn{1}{|c|}{} & \multicolumn{1}{c|}{Items} & {Rating} & {Items} & {Rating} & \multicolumn{1}{c|}{} & \multicolumn{1}{c|}{} \\
\hline
Random & \multicolumn{2}{c|}{\tabincell{l}{null}} & randomly chosen & \tabincell{l}{normal dist around \\ system mean.} & null & $r_{max}/r_{min}$ \\ \hline
Average & \multicolumn{2}{c|}{\tabincell{l}{null}} & randomly chosen & \tabincell{l}{normal dist around \\ item mean.} & null & $r_{max}/r_{min}$ \\ \hline
\tabincell{c}{Bandwagon (average)} & popular items & $r_{max}/r_{min}$ & randomly chosen & \tabincell{l}{normal dist around \\ item mean.} & null & $r_{max}/r_{min}$ \\ \hline
\tabincell{c}{Bandwagon (random)} & popular items & $r_{max}/r_{min}$ & randomly chosen & \tabincell{l}{normal dist around \\ system mean.} & null & $r_{max}/r_{min}$ \\ \hline
Segment & segmented items & $r_{max}/r_{min}$ & randomly chosen & $r_{min}/r_{max}$ & null & $r_{max}/r_{min}$ \\ \hline
Reverse Bandwagon & unpopular items & $r_{min}/r_{max}$ & randomly chosen & system mean & null & $r_{max}/r_{min}$ \\ \hline
Love/Hate & \multicolumn{2}{c|}{\tabincell{l}{null}} & randomly chosen & $r_{min}/r_{max}$ & null & $r_{max}/r_{min}$ \\ \hline
AOP & \multicolumn{2}{c|}{\tabincell{l}{null}} & \multicolumn{2}{c|}{\tabincell{l}{x-\% popular items, ratings set with \\ normal dist around item mean.}} & null & $r_{max}/r_{min}$ \\ \hline
PIA-AS & \multicolumn{2}{l|}{\tabincell{l}{power items, ratings set with normal \\ dist around item mean.}} & \multicolumn{2}{c|}{\tabincell{l}{null}} & null & $r_{max}/r_{min}$ \\ \hline
PIA-ID & \multicolumn{2}{l|}{\tabincell{l}{power items, ratings set with normal \\ dist around item mean.}} & \multicolumn{2}{c|}{\tabincell{l}{null}} & null & $r_{max}/r_{min}$ \\ \hline
PIA-NR & \multicolumn{2}{l|}{\tabincell{l}{power items, ratings set with normal \\ dist around item mean.}} & \multicolumn{2}{c|}{\tabincell{l}{null}} & null & $r_{max}/r_{min}$ \\ \hline
PUA-AS & \multicolumn{2}{l|}{\tabincell{l}{copy ratings and items from power \\ user profiles.}} & \multicolumn{2}{c|}{\tabincell{l}{null}} & null & $r_{max}/r_{min}$ \\ \hline
PUA-ID & \multicolumn{2}{l|}{\tabincell{l}{copy ratings and items from power \\ user profiles.}} & \multicolumn{2}{c|}{\tabincell{l}{null}} & null & $r_{max}/r_{min}$ \\ \hline
PUA-NR & \multicolumn{2}{l|}{\tabincell{l}{copy ratings and items from power \\ user profiles.}} & \multicolumn{2}{c|}{\tabincell{l}{null}} & null & $r_{max}/r_{min}$ \\ \hline
\end{tabular}}
\end{center}
\end{table}

3) Bandwagon (average) attack: $I_S$ contains a set of popular items. And then, we use these items as $I_S$, $\sigma(i)=r_{max} / r_{min}$ (push/nuke) and $\rho(i) \sim N(\overline{r_i},\overline{\sigma_i}^2)$ \cite{Wu2014}.

4) Bandwagon (random) attack: $I_S$ contains a set of popular items, $\sigma(i)=r_{max} / r_{min}$ (push/nuke) and $\rho(i) \sim N(\overline{r},\overline{\sigma}^2)$ \cite{Wu2014}.

5) Segment attack: $I_S$ contains a set of segmented items. And then, we use these items as $I_S$, $\sigma(i)=r_{max} / r_{min}$ (push/nuke) and $\rho(i)=r_{min} / r_{max}$ (push/nuke) \cite{Gunes2012}.

6) Reverse Bandwagon attack: $I_S$ contains a set of unpopular items, $\sigma(i)=r_{min} / r_{max}$ (push/nuke) and $\rho(i) \sim N( \overline{r}, \overline{\sigma}^2)$ \cite{Gunes2012}.

7) Love/Hate attack: $I_S=\phi$ and $\rho(i)=r_{min} / r_{max}$ (push/nuke) \cite{Gunes2012}.

8) AOP attack: A simple and effective strategy to obfuscate the Average attack is to choose filler items with equal probability from the top x\% of most popular items rather than from the entire collection of items \cite{Seminario2014b}.

9) PIA-AS attack: The top-N items with the highest aggregate similarity (AS) scores become the selected set of power items. This method requires at least 5 users who have rated the same item $i$ and item $j$  \cite{Seminario2014b}.

10) PIA-ID attack: Based on In-Degree centrality, power items participate in the highest number of similarity neighborhoods. For each item $i$ compute similarity with every item $j$ applying significance weighting $\frac{n_{cij}}{50}$, where $n_{cij}$ is the number of users that have rated the same items $i$ and $j$, then discard all but the top-N neighbors for each item $i$. Count the number of similarity scores for each item $j$, and select the top-N item $j$'s \cite{Seminario2014b}.

11) PIA-NR attack: Power items are the items with the highest number of user ratings. We select the top-N items based on the total number of user ratings they have in their profile \cite{Seminario2014b}.

12) PUA-AS attack: The top 50 users with the highest Aggregate Similarity scores become the selected set of power users. This method requires at least 5 co-rated items between user $u$ and user $v$ and does not use significance weighting \cite{Seminario2014a}.

13) PUA-ID attack: Based on the In-Degree centrality concept from social network analysis, power users are those who participate in the highest number of neighborhoods. For each user $u$ compute its similarity with every other user $v$ applying significance weighting, then discard all but the top 50 neighbors for each user $u$. Count the number of similarity scores for each user $v$ and select the top 50 user $v$'s \cite{Seminario2014a}.

14) PUA-NR attack: Power users are the users with the highest number of ratings. We selected the top 50 users based on the total number of ratings they have in their user profile \cite{Seminario2014a}.

\section{Our approach}

In this section, we first present an overall introduction of our approach. Then, the two main aspects of our work including features extraction from user profiles and
RAdaBoost for attack detection are described in detail.

\subsection{The framework of our approach}

As shown in Figure 1, our approach consists of four phases: the phase of constructing training dataset and test datasets, the phase of feature extraction, the phase of training classifier via RAdaBoost, and the phase of test for generating detection results.
At the phase of constructing training set and test sets, the data sets are constructed by attack profiles (diverse attack models are injected) and genuine profiles.
Concretely, for training data set, we use several representative attack models such as Random, Average attacks etc. to generate mixed attack profiles. Specially, we modest increase
the number of attacks (160 attackers for each attack models) when constructing the training data set aim to relieve the extent of imbalance in training phase
(more details in section 5). Then, we combine them with genuine profiles as the our training data set. For test data sets, attack profiles  with different filler sizes and attack sizes are inserted into the genuine profiles to form the test data sets (see section 5).
At the phase of feature extraction, we employ 18 features (more details in the next subsection) extracted from user profiles to characterize a feature representation (or feature vector) for each user in both training data set and test data sets.
At the phase of training, we use RAdaBoost to train a  strong composite estimator (or classifier) based on training features.
Finally, we use features retrieved from test data sets as the input into the obtained trained estimator and generate detection results at the phase of testing.

\subsection{Feature extraction from user profiles}
%There are different metrics that have been proposed to measure the features of user profiles. Our aim is to learn to label each profile as either being part of an attack or as coming from a genuine user by extracting features from user profiles. All features used in this paper are summarized in Table 2. These features come in three varieties: generic features, features based on the filler size  and model-specific features. The generic features are basic descriptive statistics attempt to capture characteristics that tend to make an attacker's profile look different from a genuine user \cite{Burke2006}, such as RDMA, WDMA and etc. The features based on the filler size can characterize the distribution features of different items \cite{Zhang2014}, such as FSTI et al. For the model-specific features, all features are implemented to detect characteristics of profiles generated by the specific attack models \cite{Burke2006}, such as FMTD et al.
%{\bf describe features more detailed. Sectionalized description maybe better }
%\cite{Williams2007a}
Previous works \cite{Burke2006, Williams2007a, Williams2007b, Morid2014} summarized different metrics to characterize the features extracted from user profiles. These features generally fall into two types: generic and type-specific features.
The generic features are basic descriptive statistics that attempt to discriminate between attack profiles and genuine profiles and the type-specific features are implemented to detect characteristics of profiles generated by specific attack models or specific signatures of attacks.
In the present work, we employ 10 features from these two types. Besides, we also employ 5 features based on the filler size \cite{Zhang2014} and propose additional 3 new features which measure the distribution of specific rating such as mean rating, maximum rating and minimum rating in filler items for each user.
\subsubsection{Generic features}
Attack profiles usually have high deviation from the mean value for the target items and low deviation from the mean value for remaining items. Thus, generic features such as RDMA, WDMA etc. are often used to measure the deviation of rating for user profiles \cite{Burke2006, Williams2007a, Williams2007b, Zhang2014b}.

\begin{figure}[H]
%\addtolength{\tabcolsep}{-3pt}
\centering
\subfigure{\label{Fig.sub.a}\includegraphics[height=5cm,width=12cm]{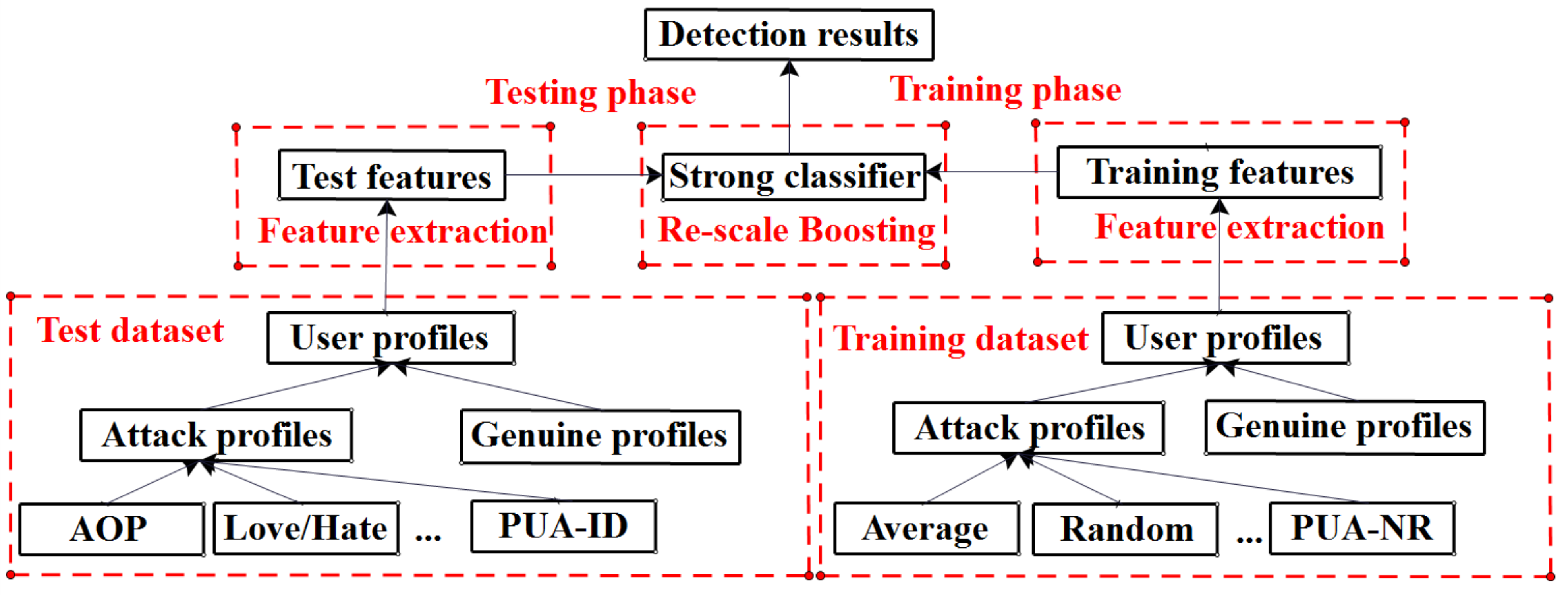}}
\caption{The framework of our approach.}
\label{f1}
\end{figure}

Rating Deviation from Mean Agreement (RDMA):
\begin{eqnarray}
RDMA_u=\frac{\sum_{i=0}^{N_u}\frac{ \left | r_{u,i}-\overline{r_i} \right | }{NR_i}}{N_u}
\end{eqnarray}
where $N_u$ is the number of ratings that user $u$ has rated and $NR_i$ is the number of ratings provided for item $i$. $r_{u,i}$ denotes the rating given by user $u$ to item $i$, $\overline{r_i}$ denotes the mean rating of item $i$ across all users.

 Weighted Deviation from Mean Agreement (WDMA):
\begin{eqnarray}
WDMA_u=\frac{\sum_{i=0}^{N_u}\frac{ \left | r_{u,i}-\overline{r_i} \right | }{NR_i^2}}{N_u}
\end{eqnarray}

Weighted Degree of Agreement (WDA):
\begin{eqnarray}
WDA_u=\sum_{i=0}^{N_u}\frac{ \left | r_{u,i}-\overline{r_i} \right | }{NR_i}
\end{eqnarray}

Length Variance (LengthVar):
\begin{eqnarray}
LengthVar_u=\frac{\left | n_u - \overline{n} \right | } {\sum_{k \in U} {(n_k - \overline{n})}^2 }
\end{eqnarray}
where $n_u$ is the total number of ratings in the system for user u. $U$ is the total number of users in the system. $\overline{n}$ is the average length of a profile in the system.

\subsubsection{Type-specific features}
Model-based methods assume that we have some prior knowledge about the attack models. Based on an assumed model, ratings can be automatically divided into filler items and selected items \cite{Burke2006, Williams2007a, Williams2007b, Zhang2014b}. Therefore, the measurements such as MeanVar, FMTD etc. can be calculated from each subset to measure the authenticity of profiles.

Mean Variance (MeanVar):
\begin{eqnarray}
MeanVar_u=\frac{\sum_{j \in {P_{u,F}}}{(r_{u,j}-\overline{r_u})}^2}{\left | P_{u,F} \right | }
\end{eqnarray}
where $P_{u,F}$ is the rest of the profile: $P_u-P_{u,T}$, $P_{u,T}=\{ i \in P_u, such\ that\ r_{u,i}=r_{max} \}$ (or $r_{min}$ for nuke attack), $P_u$ is the profile of user $u$.

 Filler Mean Target Difference (FMTD):
\begin{eqnarray}
FMTD_u=\left | \frac{\sum_{i \in P_{u,T}} r_{u,i}  } {\left | P_{u,T} \right | } - \frac{\sum_{k \in P_{u,F}} r_{u,k} } {\left | P_{u,F} \right | } \right |
\end{eqnarray}

 Target Model Focus (TMF):
\begin{eqnarray}
TMF_u=\max_{j \in P_T} F_j
\end{eqnarray}
where $F_i= ({\sum_{u \in U} \phi_{u,i}}) / ({\sum_{u \in U} \left | P_{u,T} \right |} )$, and $\phi_{u,i}$ is 1 if $i \in P_{u, T}$, 0 otherwise. $P_T$ denotes the item set of potential targets \cite{Williams2007b}.

 Filler Mean Variance (FMV):
\begin{eqnarray}
FMV_u=\frac{1}{\left | U_u^{F_m} \right | } \sum_{i \in U_u^{F_m}} {(r_{u,i}-\overline{r_i} )}^2
\end{eqnarray}
where $U_u^{F_m}$ is the partition of the profile of user $u$ hypothesized to be the set of filler items $F$ by model $m$. $\left | U_u^{F_m} \right |$ is the number of items in the hypothesized filler partition of profile $P_u$ by model $m$.

Filler Mean Difference (FMD):
\begin{eqnarray}
FMD_u=\frac{1}{\left | U_u \right |}\sum_{i=1}^{\left | U_u \right |} \left | r_{u,i}-\overline{r_i} \right |
\end{eqnarray}
where $U_u$ is the partition of the profiles of user $u$. $\left | U_u \right |$ is the number of the profiles of user $u$.

 Filler Average Correlation (FAC):
\begin{eqnarray}
FAC_u=\frac{\sum_{i \in {I_u} }{(r_{u,i} - \overline{r_i})} }  { \sqrt{ \sum_{i \in {I_u} }{(r_{u,i}- \overline{r_i} )}^2 } }
\end{eqnarray}
where $I_u$ is the set of items rated by user $u$.

\subsubsection{Features based on the filler size}

User profiles with different number of ratings will generate different features. Similarly, the number of rating on different types of items will also generate different features. Such as FSTI, FSPI etc. \cite{Zhang2014}.

Filler Size with Total Items (FSTI):
The ratio between the number  of items rated by user $u$ and the number of entire  items in the recommender  system \cite{Zhang2014}.
\begin{eqnarray}
FSTI_u=\frac{\sum_{i=1}^{ \left | I \right |} O{(r_{u,i})} } {\left | I \right | }
\end{eqnarray}
where $I$ is the set of items in the system. $\left | I \right |$ denotes the total number of items in the system. $O{(r_{u,i})}$ is 1 if user $u$ rated item $i$, 0 otherwise.

Filler Size with Popular Items (FSPI):
The ratio between the number of popular items rated by user $u$ and the number of entire popular items in the recommender system \cite{Zhang2014}.
\begin{eqnarray}
FSPI_u=\frac{\sum_{i=1}^K {O(r_{u,i})} }{ K }
\end{eqnarray}
where $K$ denotes the boundary point of popular items and unpopular items.

Filler Size with Popular Items in Itself (FSPII):
The ratio between the number of popular items rated by user $u$ and the number of entire items rated by user $u$ \cite{Zhang2014}.
\begin{eqnarray}
FSPII_u=\frac{\sum_{i=1}^K {O(r_{u,i})} }{ \sum_{j=1}^{ \left | I \right |} O{(r_{u,j})} }
\end{eqnarray}

Filler Size with Unpopular Items (FSUI):
The ratio between the number of unpopular items rated by user $u$ and the number of entire unpopular items in the recommender system \cite{Zhang2014}.
\begin{eqnarray}
FSUI_u=\frac{\sum_{i=1}^{ \left | I \right |} O{(r_{u,i})} } { \left | I \right | - K }
\end{eqnarray}

Filler Size with Unpopular Items in Itself (FSUII):
The ratio between the number of unpopular items rated by user $u$ and the number of entire items rated by user $u$ \cite{Zhang2014}.
\begin{eqnarray}
FSUII_u=\frac{\sum_{i=K+1}^{ \left | I \right |} O{(r_{u,i})} } { \sum_{k=1}^{ \left | I \right |} O{(r_{u,k})}  }
\end{eqnarray}

\subsubsection{Our proposed features}
We propose 3 new features which focus on the number of specific ratings (such as the maximum score, minimum score and average score) on filler or selected items. Since attackers show different attack intents in CFRSs, the filler or selected set of attack profiles may be filled by specific items (i.e., select popular items for Bandwagon (average and random) attacks, select randomly items in the system for Random attack) with the highest score or the lowest score or average score. Take nuke attacks for example, the selected items or filler items are rated with maximum score in Reverse Bandwagon, Segment and Love/Hate  attacks (as shown in Table 2). Similarly, the selected items or filler items are rated with minimum score in Bandwagon (average), Bandwagon (random) and Segment attacks. In Random attack, the filler items are rated with some average score (normal distribution around system mean).
Therefore, the number of specific ratings can be used to evaluate partly the difference between genuine profiles and attack profiles.

Filler Size with Maximum Rating in Itself (FSMAXRI):
The ratio between the number of items rated by user $u$ with maximum score and the number of entire items rated by user $u$.
\begin{eqnarray}
FSMAXRI_u=\frac{\sum_{i=1}^{I_u} O{(r_{u,i} = r_{max})} } { \sum_{k=1}^{I_u} O{(r_{u,k})}  }
\end{eqnarray}
where $r_{u,i}$ is the rating given by user $u$ to item $i$, $r_{max}$ is the maximum score in the system. $I_u$ denotes the set of items rated by user $u$. $O{(r_{u,i} = r_{max})}$ is 1 if user $u$ rated item $i$ with rating $r_{max}$, 0 otherwise. $O{(r_{u,k})}$ is 1 if user $u$ rated item $k$, 0 otherwise.

Filler Size with Minimum Rating in Itself (FSMINRI):
The ratio between the number of items rated by user $u$ with minimum  score and the number of entire items rated by user $u$.
\begin{eqnarray}
FSMINRI_u=\frac{\sum_{i=1}^{ \left | I \right |} O{(r_{u,i} = r_{min})} } { \sum_{k=1}^{ \left | I \right |} O{(r_{u,k})}  }
\end{eqnarray}
where $r_{min}$ is the minimum score in the system. $O{(r_{u,i} = r_{min})}$ is 1 if user $u$ rated item $i$ with rating $r_{min}$, 0 otherwise.

Filler Size with Average Rating in Itself (FSARI):
The ratio between the number of items rated by user $u$ with average score and the number of entire items rated by user $u$.
\begin{eqnarray}
FSARI_u=\frac{\sum_{i=1}^{ \left | I \right |} O{(r_{u,i} = r_{avg})} } { \sum_{k=1}^{ \left | I \right |} O{(r_{u,k})}  }
\end{eqnarray}
where $r_{avg}$ is the average score in the system. $O{(r_{u,i} = r_{avg})}$ is 1 if user $u$ rated item $i$ with rating $r_{avg}$, 0 otherwise.

\subsection{Re-scale AdaBoost for attack detection}
After the raw user profiles are transformed to a set of sophisticated features, an effective detection method based on these features for ``shilling'' attacks is crucial. As is known,
the number of attackers is usually far smaller than genuine users in CFRSs, thus
the supervised learning based attack detection can be formulated as an imbalanced classification, actually.
Conventional supervised learning based detection method (i.e., SVM or kNN) often inevitably have individual weaknesses for  handling this kind of issues. Under this circumstance, %ensemble learning that combines multiple algorithms (or weak learners) together comes into our sights.
%Ensemble techniques such as the Bagging \cite{Breiman1996}, Boosting \cite{Freund1995}, Stacking \cite{Smyth1999}, Bayesian  averaging \cite{Mackay1991} and Random forest \cite{Breiman2001} can take the form of using different algorithms, using the same algorithm with
%different settings, or assigning different parts of the data set to different classifiers in practice and benefit from favorable learning capability.
%Typically,  Boosting  is one of the most popular meta-algorithms paradigm for supervised learning based on any fitting criterion and its variants are based on a rich theoretical analysis and generalization capability verification in \cite{Bagirov2010}, \cite{Bartlett2007}, \cite{Bickel2006}, \cite{Buhlmann2003}, \cite{Friendman2000}, \cite{Lin2013}, \cite{Lin2015}, \cite{Zhang2005}.
%Different from other ensemble methods that build the merge results from multiple  weak learners, Boosting is unique because it is sequential, i.e., it is a forward stagewise procedure.
Boosting comes into our sights as it has been proved to be efficient when faced with some difficult scenarios as imbalanced classification \cite{Harrington2012}.
In Boosting, weak learners are fitted iteratively to the training data, using appropriate methods to gradually increase emphasis on observations modelled poorly by the existing collection of weak learners.  More specifically, AdaBoost apply {\bf weights} to the observations (or samples), emphasising poorly modelled ones and gradually (or iteratively, more precisely) strengthening the correction of misclassifications.
The following Algorithm \ref{alg1} interpret the main idea of AdaBoost \cite{Freund1995a}.

\begin{algorithm}[H]\caption{\bf AdaBoost}\label{alg1}
\begin{algorithmic}
\STATE {{ Step 1: (\bf Initialization)}: Given data
$\{(x_i,y_i):i=1,\dots,m\}$, where $x \in R^d$ and $y \in \{-1,+1\}$, weights $\{(w^{(1)}_i)= \frac{1}{m}:i=1,\dots,m\}$
, dictionary $\mathcal D_n=\{g_1,\dots,g_n\}$, iteration number $T$
and $f_0\in\mbox{span}(\mathcal D_n)$}.
\STATE { Step 2: Find $g_t\in \mathcal D_n$ such that
 minimizes the weighted sum error $$ \epsilon_t  = Pr_{i \thicksim w^{(t)} }[{g_t}({x_i}) \ne {y_i}] = \sum\limits_{i = 1}^m {{w_i}^{(t)}1_{({g_t}({x_i}) \ne {y_i})}}  $$ for misclassified samples.
 }
 \STATE{ Step 3: Choose $$\alpha_t = \frac{1}{2} \ln \left(\frac{1-\epsilon_t}{\epsilon_t}\right)$$ and update weights
$$ w^{(t)}_{i} = \frac {w^{(t-1)}_{i} exp^{-y_i \alpha_t g_t(x_i)}} {Z_t} $$ for all samples,
where $Z_t=2[\epsilon_t(1-\epsilon_t)]^{1/2}$ is a normalization factor.}

 \STATE{ Step 4: Add to ensemble
 $f_{t} = f_{t-1} + \alpha_t g_t$.

}
 \STATE{  Step 5: Increase $t$ by one and repeat Step 2 and Step 3 if
$t<T$.}
\end{algorithmic}
\end{algorithm}

%\begin{remark}
%In the step 3 in Algorithm \ref{alg1}, it is easy to check that if the weak learners satisfy $\{ \|g_i\|_m=1, \ i=1, \dots, n \}$, then
%$$
%            \beta_k= \arg\min_{\beta_k \in
%            \mathbf R}\mathcal E_{\bf {z}}(f_{k-1}+\beta_k g_k)= \langle r_{k-1},g_k\rangle_m.
%$$
%Therefore, we call it as the linear search step and $\beta_k=\langle r_{k-1},g_k\rangle_m$ as the step-size. And it is easy to show that if $\beta_k= ln(\sqrt {\frac{{1 - {\varepsilon _k}}}{{{\varepsilon _k}}}} )$, where $\varepsilon _k$ is the re-weighted error applied in AdaBoost.
%Then the AdaBoost algorithm in  \cite{Freund1996} can be formulated as  above Algorithm \ref{alg1}.
%\end{remark}
From a statistical view, AdaBoost also can be viewed as a form of "Gradient Boosting Machine" \cite{Friendman2001}. Consider a loss
function in this case, a measure that represents the loss in predictive performance due to a sub-optimal model. Boosting  is a numerical optimisation technique for minimising the loss function by adding at each step a new weak learner that best reduces (steps down the gradient of) the loss function.
 Original gradient Boosting algorithm was proved to be consistent, which can be easily deduced by applying the method in \cite{Barron2008} to \cite[Theorem1]{Livshits2009}, however, a number of studies \cite{Devore1996,Livshits2009,Temlyakov2008a} also showed that its approximation  rate is far slower. The numerical
convergence rate of Boosting lies in $(C_0t^{-0.1898},C_0't^{-0.182})$, which is much slower than the
minimax nonlinear approximation rate $\mathcal O(t^{-1/2})$. Here
and hereafter, $t$ denotes the number of iterations, and $C_0,C_0'$
are absolute constants.

Recently, Lin et al. \cite{Lin2015} and Xu et al. \cite{Xu2015b} proposed a  re-scale Boosting (RBoosting) to improve  the performance of original gradient Boosting.
Different from the aforementioned strategies that focus on controlling  the step-size of  $g_t^*$ such as some existing variants like Regularized shrinkage Boosting \cite{Ehrlinger2012}, Regularized truncated Boosting \cite{Zhang2005}, $\varepsilon$-Boosting \cite{Hastie2007}, they cheered  a novel direction
to improve the numerical convergence rate and consequently, the
generalization capability of Boosting. The core idea is that if the
approximation (or learning) effect of the  $t$th iteration is not
good, then we regard  $f_t$ to be too aggressive and therefore shrink it within a certain extent.
By such an interesting modification, the optimal numerical convergence of RBoosting can be guaranteed. This means that, RBoosting is among the almost optimal nonlinear approximant and therefore, RBoosting may possess better learning performance than other Boosting-type algorithms.
Based on the general idea of RBoosting, the re-scale AdaBoost (RAdaBoost) can be interpreted
as the following Algorithm 2.

\begin{algorithm}[H]\caption{\bf  Re-scale AdaBoost}\label{alg2}
\begin{algorithmic}
\STATE {{ Step 1: (\bf Initialization)}: Given data
$\{(x_i,y_i):i=1,\dots,m\}$, where $x \in R^d$ and $y \in \{-1,+1\}$, weights $\{(w^{(1)}_i)= \frac{1}{m}:i=1,\dots,m\}$
, dictionary $\mathcal D_n=\{g_1,\dots,g_n\}$, a set of shrinkage
degree $\{s_t\}^{t^*}_{t=1}$ where $s_t=2/(t+u), u \in \mathbf{N}$, iteration number $T$
and $f_0\in\mbox{span}(\mathcal D_n)$}.
\STATE { Step 2: Find $g_t\in \mathcal D_n$ such that
 minimizes the weighted sum error $$ \epsilon_t  = Pr_{i \thicksim w^{(t)} }[{g_t}({x_i}) \ne {y_i}] = \sum\limits_{i = 1}^m {{w_i}^{(t)}1_{({g_t}({x_i}) \ne {y_i})}}  $$ for misclassified samples.
 }
 \STATE{ Step 3: Choose $$\alpha_t = \frac{1}{2} \ln \left(\frac{1-\epsilon_t}{\epsilon_t}\right)$$ and update weights
$$ w^{(t)}_{i} = \frac {w^{(t-1)}_{i} exp^{-y_i \alpha_t g_t(x_i)}} {Z_t} $$ for all samples,
where $Z_t=2[\epsilon_t(1-\epsilon_t)]^{1/2}$ is a normalization factor.}

 \STATE{ Step 4: Add to ensemble
 $f_{t} = (1-s_t)f_{t-1} + \alpha_t g_t$.

}
 \STATE{  Step 5: Increase $t$ by one and repeat Step 2 and Step 3 if
$t<T$.}
\end{algorithmic}
\end{algorithm}

%\noindent

\section{Experiments and analysis}
In this part,  we firstly introduce the
experimental settings, including the data sets, evaluation metrics and
computational environment. Secondly, the impact of the extracted features are analyzed.
Then, we compare the performance of RAdaBoost with three other
benchmarked methods such as SVM, kNN and AdaBoost on diverse 4 attack detection methods
to demonstrate the outperformance of RAdaBoost. Finally,  the remaining 10 types of attacks are conducted by means of RAdaBoost to further evaluate its performance.

%\subsection{Experimental settings}

%\subsection{Normal datasets}
\subsection{Experimental settings}
In our experiments, we use the MovieLens-100K \footnote{http://grouplens.org/datasets/movielens/} dataset as the data set describing the behaviors of genuine users in recommender system. MovieLens-100K was collected by the GroupLens Research Project at the University of Minnesota. It is the one of the most popular data sets used by researchers and developers in the field of collaborative filtering and attack detection in recommender systems.
It consists of 100,000 ratings on 1682 movies by 943 raters and each rater had to rate at least 20 movies. All ratings are in the form of integral values between minimum value 1 and maximum value 5. The minimum score means the rater distastes the movie, while the maximum score means the rater enjoyed the movie. According to the information derived from MovieLens website, the sparse ratio \footnote{The ratio between the number of ratings and entire ratings in the rating matrix.} of the rating data approximates to 93.7\% and the average rating of all users is around 3.53. Besides, the Average Number of Items Rated (ANIR) by each user is approximately 7\%.
Attack profiles are generated according to different attack models (as shown in Table 2).
The attack profiles indicate the attacker¡¯s intention that he wishes a particular item can be rated the highest or lowest rating.
In this paper, we just detect the nuke attacks and the push attacks can be detected in the analogous manner.
For each attack model, we generate nuke attack profiles according to the corresponding attack models with different attack sizes \{1.1\%, 6.4\%, 11.7\%, 17.0\%, 22.3\%, 27.6\%\} and filler sizes \{1.2\%, 4.2\%, 7.3\%, 10.3\%, 13.3\%, 16.4\%\}.
To ensure the rationality of the results, the target item is randomly selected for each attack profile.
In addition, for bandwagon attacks, we select movies \{50, 56, 100, 127, 174, 181, 258, 286, 288, 294\} as the popular movies which are rated by more than 300 users in the system. In segment attack, we use movies \{50, 183, 185, 200, 234, 443\} as the segmented movies \cite{Li2011}. And for Reverse Bandwagon attack, we randomly choose 10 movies as the selected movies which are rated by one user in the system.
For training set, we use the whole MovieLens-100K dataset to generate a  attack profiles by exploiting 7 representative known attack models (random, average, bandwagon (average), segment, reverseBandwagon, PIA-ID and PUA-NR) with 17.0\% attack size (160 attackers) and diverse filler sizes \{1.2\%, 4.2\%, 7.3\%, 10.3\%, 13.3\%, 16.4\%\}. And then, we combine these 7 attack datasets into MovieLens-100K dataset to construct a mixed user profiles as our training data. Thus, the training dataset consists of 943 genuine users and 1120 ($160\times7$) attackers.
For test data sets, based on the whole MovieLens-100K dataset, we generate respectively attack profiles by exploiting 14 attack models with different attack sizes \{1.1\%, 6.4\%, 11.7\%, 17.0\%, 22.3\%, 27.6\%\} and filler sizes \{1.2\%, 4.2\%, 7.3\%, 10.3\%, 13.3\%, 16.4\%\}. And then, the generated attack profiles are respectively inserted into genuine profiles to construct our test datasets. Therefore, we have 504 ($14\times6\times6$) test datasets including 14 attack models, 6 different attack sizes and 6 different filler sizes.

To measure the effectiveness of the proposed detection methods, we use three metrics such as classification error, detection rate and false alarm rate in the test sets, which are used in similar experiments \cite{Chung2013}.
Classification error is defined
as the number of misclassifications divided by the number of
all test user profiles.
\begin{eqnarray}
classification\ error= \frac{{{\rm{\#  Misclassifications}}}}{{{\rm{\# User\ Profiles}}}}
\end{eqnarray}
Detection rate is defined
as the number of detected attack profiles divided by the number of
attack profiles.
\begin{eqnarray}
detection\ rate= \frac{{{\rm{\#  Detection}}}}{{{\rm{\#  Attack\ Profiles}}}}
\end{eqnarray}
False alarm rate is the number of genuine profiles that
are predicted as attack profiles divided by the number of genuine
profiles.

\begin{eqnarray}
false alarm\ rate= \frac{{{\rm{\#  False\ alarm}}}}{{{\rm{\#  Genuine\ Profiles}}}}
\end{eqnarray}

All numerical studies are implemented using MATLAB R2014a on a Windows
personal computer with Core(TM) i7-3770 3.40GHz CPUs and RAM 16.00GB.

\subsection{Impact of extracted features}

%For nuke attacks, FSMAXRI and FSMINRI (see subsection 4.2) can effectively characterize Bandwagon (average), Segment and Reverse Bandwagon attack profiles. The rating that rated on filler items with system mean (the average rating in the system) or item mean (the average rating in each user profile) can be evaluated by FSARI, such as in Bandwagon (average) and Random attacks.

%with the filler size increasing, the false alarm rate of 18 features based is significantly lower than the others in the late stage (filer size $>$ 4.2\%) under all attack models except for PIA-AS attack. It is worth noting that the false alarm rate of 18 features based also shows lower than others in the intermediate stage (1.2\% $<$ filler size $<$ 13.3\%) for PIA-AS attack.
%
% These results indicate that the added 3 features are useful to further discriminate between attacker and genuine users with the filler size increased.

 %The goal is to expect that the 3 new features of our proposed can further improve the detection performance in comparison with the 10 features based and 15 features based.  

\begin{figure}[H]
%\addtolength{\tabcolsep}{-3pt}
\centering
\subfigure{\label{Fig.sub.a}\includegraphics[height=8.1cm,width=13cm]{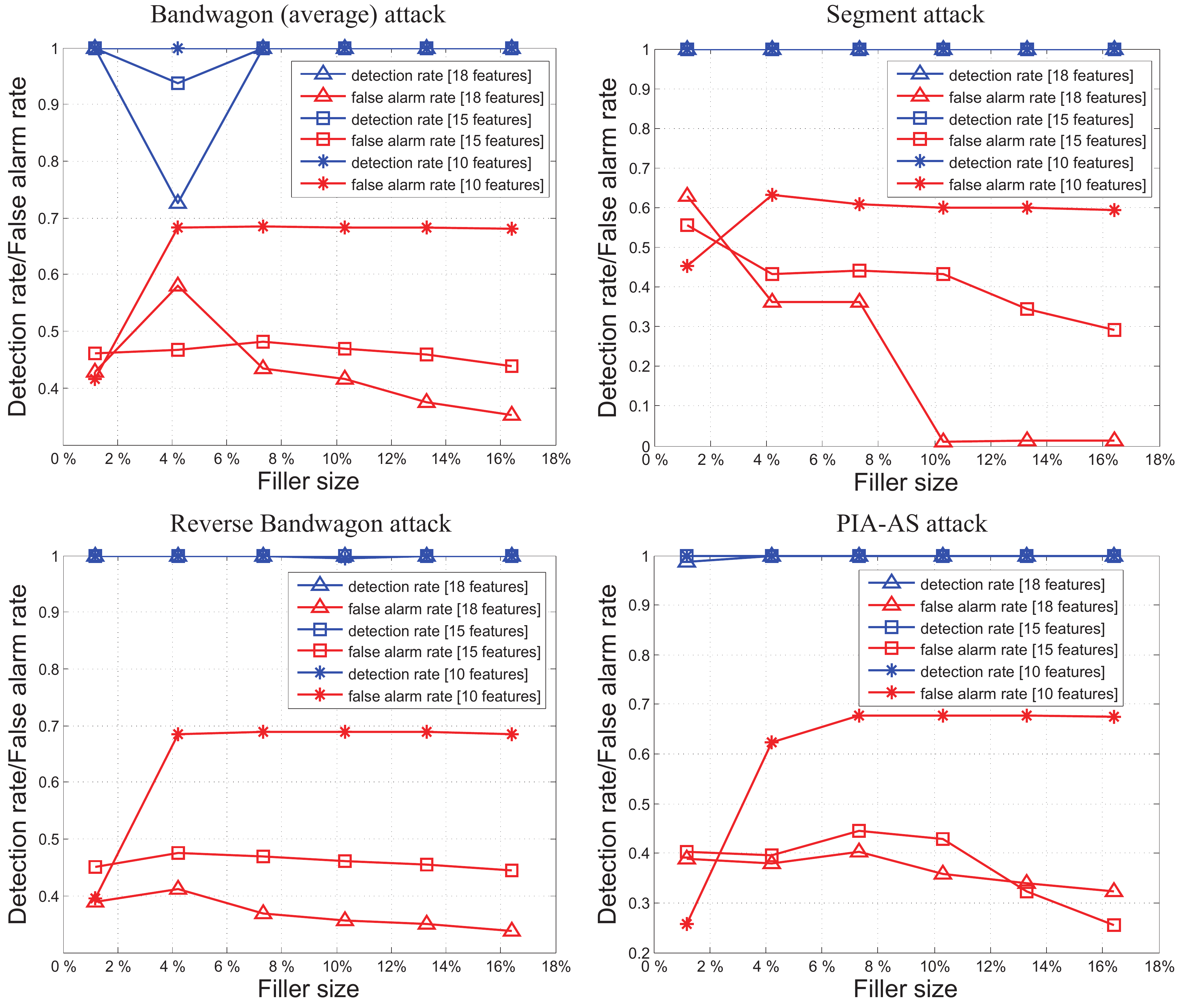}}
\caption{Relationship between the number of extracted features and the performance with respect to the filler size. Attack size is 17.0\%.}
\label{f1}
\end{figure}

\begin{figure}[H]
%\addtolength{\tabcolsep}{-3pt}
\centering
\subfigure{\label{Fig.sub.a}\includegraphics[height=6.5cm,width=11cm]{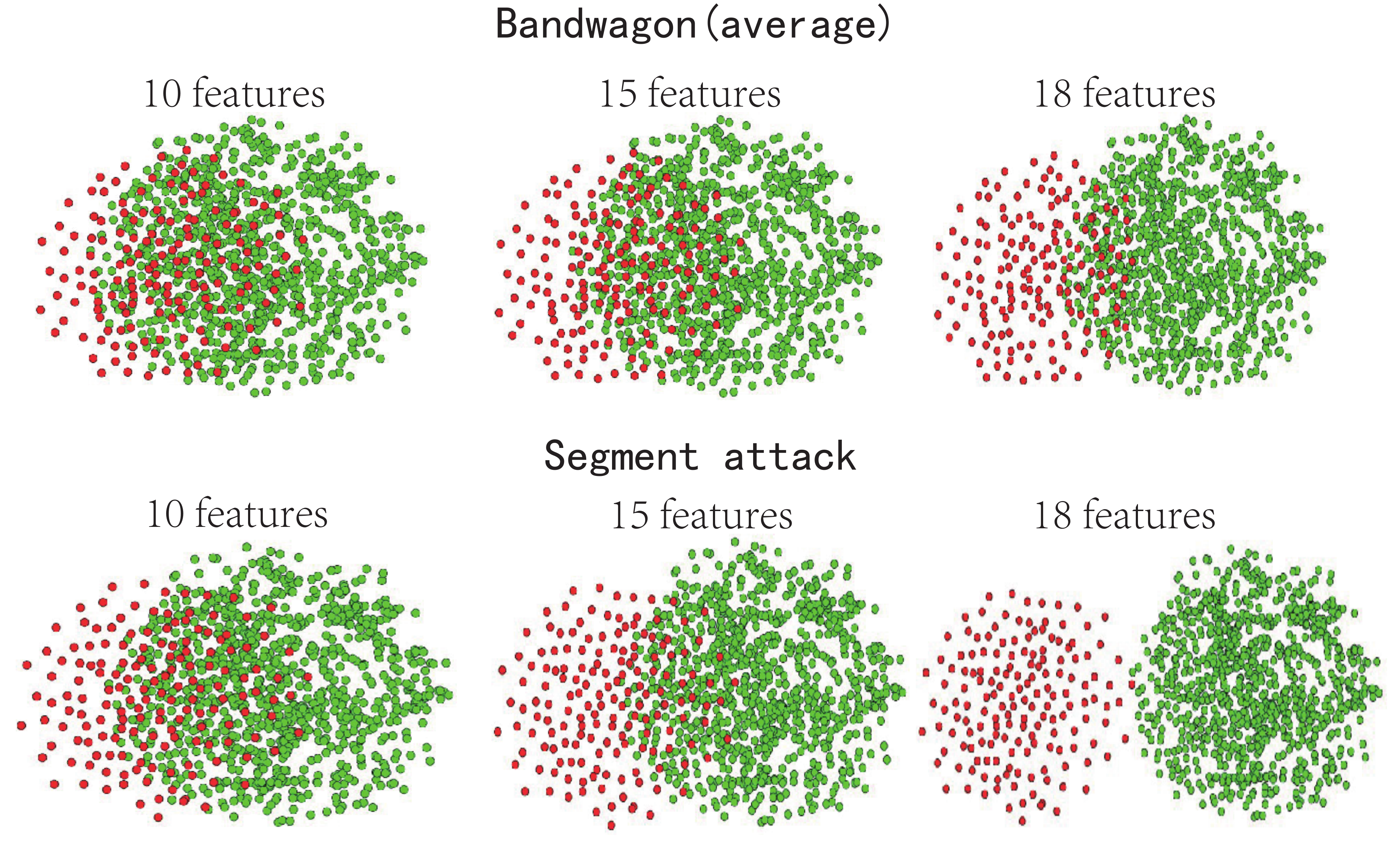}}
\caption{The diagrams of clustering results with diverse feature employed, where red nodes denote attackers and green nodes denote genuine users.}
\label{f1}
\end{figure}

To evaluate the impact of the extracted features, we conduct a list of experiments in several attack models with diverse filler sizes as Figure 2 illustrated.
We utilize EM (Expectation-maximization) clustering method (Clustering results and EM clustering method were created using Weka \footnote{http://www.cs.waikato.ac.nz/ml/weka/}) to separate attackers from genuine users as far as possible based on 10 features (generic and type-specific features), 15 features (additional 5 features based on filler size) and 18 features (all aforementioned features including 3 our proposed features), respectively, in order to analyze the relationship between the number of extracted features and the performance with respect to the filler size. Just as shown in Figure 2, Bandwagon (average), Segment, Reverse Bandwagon and PIA-AS attacks are taken for examples.
It is distinctly observed from the results that the false alarm rate significantly decrease with using more extracted features.
Furthermore, we take two diagrams to intuitively show the clustering results as shown in Figure 3 (Bandwagon (average) and Segment attacks are taken for examples). By fixing attack size (17.0\%, 160 attackers) and filler size (13.3\%, 170 items), the strikingly clustering results illustrate that hard classification task becomes easier to perform with more
well-designed features employed.

\subsection{Experimental results and analysis}

First, we compare the detection performance of RAdaBoost with three benchmarked methods such as SVM, kNN and AdaBoost on 4 test attack profiles described above to validate the outperformance of RAdaBoost. The details of setting of each method is described as follows:

 $\bullet$ SVM: LibSVM and the default parameters are employed as \cite{Chang2011} for training binary profile classifier with $Prediction = +1$ if classified as authentic and $Prediction = -1$ if classified as attack. To classify unseen test data sets, the trained SVM  model (or classifier) in the training set are used to determine the class label.

 $\bullet$ kNN: Standard kNN algorithm  is used as \cite{Hastie2001}. The $k$ nearest neighbors ($k$ is chosen by 5-folds cross validation on the training data set) in  the training set are collected for prediction using one over Pearson correlation distance weighting.

 $\bullet$ AdaBoost: We utilize  decision stumps (with the number of splits $J=1$) to build up the week learners  for classification. The number of iterations (or the number of stumps to be fitted) is also selected via 5-folds cross validation on the training data set.

 $\bullet$ RAdaBoost:   For  additional shrinkage degree parameter,
$s_k=2/(k+u), u \in \mathbf{N}$, in RBoosting, we create 20 equally spaced values of $u$ in logarithmic space between $1$ to $10^6$ and select the appropriate $u*$ as \cite{Xu2015b}. The other settings are the same as Boosting.

Fig.4-Fig.7 illustrate the performance surfaces of the RAdaBoost classifier for the aforementioned test sets, which contain 4 attack models (take AOP, Bandwagon (random), PIA-NR and PUA-NR attacks for examples) with different attack sizes \{1.1\%, 6.4\%, 11.7\%, 17.0\%, 22.3\%, 27.6\%\} and filler sizes \{1.2\%, 4.2\%, 7.3\%, 10.3\%, 13.3\%, 16.4\%\}. For comparison, the performance surfaces of SVM, kNN and AdaBoost are also presented.
It can be easily observed from these figures that,  the classification error of SVM rise with the increase of the attack size, which implies
the SVM classifier could not effectively classify and detect attacks generated by these 4 types of attack models when the total number of attacks are far smaller than genuine users.
Although SVM can achieve fairly high classification performance within some small
attack size areas, the detection rates of SVM are also small.
It shows that the high prediction accuracy is almost produced by abundant genuine users but fail to capture the little concerned attackers.
In our 4 sets of experiments, only a few bandwagon (random) attack profiles could be detected by SVM and naturally SVM barely false alarmed.
kNN essentially outperforms SVM in our 4 types of attack detection methods with lower classification error, much more higher detection rate and pimping false alarm rate. However, we also notice that the classification performance of kNN is still poor and it may fail for detection within  some certain attack and filler size areas. Just as figures showed, kNN fail to detect AOP, PIA-NR attacks when the filler size is too small and bandwagon (random) attacks when the attack size is small. And for PUA-NR attack, kNN is just slightly better than SVM and barely detected. For AOP attack detected by kNN, the results may indicate that some genuine profiles are misclassified as attack profiles since a large number of genuine profiles have the same or similar number of popular items as the AOP attack profiles when filler size is too big and small.
Compare with SVM and kNN, Boosting significantly improve the classification performance owing to it iteratively strengthen the correction of the misclassifications. And hence, AdaBoost further enhance detection rate with very low false alarm rate. However, just as figures shown, although AdaBoost can effectively detect attack over a wide range of attack and filler size, we also observe that its failure within some certain areas (i.e., AOP and PIA-NR attacks with small attack size and high filler size, bandwagon (random) attacks with small attack  and  filler size). Especially, AdaBoost can not effectively detect PUA-NR attacks  within a large high attack size area.
 As figures shown, RAdaBoost additionally improve the classification performance of AdaBoost by imposing a re-scale operator and consequence enhance detection rate with negligible false alarm rate in  the 4 types of attacks.
So far, all the  comparative experimental results illustrate that the RAdaBoost outperforms Boosting and conventional supervised learning based detection methods including SVM and kNN.
Finally, to further evaluate the effectiveness of RAdaBoost, we also conduct other 10 types of attacks to show the performance surfaces of RAdaBoost just as  Fig.8 illustrated. From results, we can distinctly observe that, except for PUA-AS and PUA-ID attacks,  RAdaBoost can effectively detect all of the attacks with almost no false alarm.
Although it also shows low detection rates for some attacks with small attack and filler size.
Comparing with previous research results \cite{Burke2006, Williams2007b, Zhang2012}, the detection performance of RAdaBoost is more optimistic.
For PUA-AS and PUA-ID attacks, which are recently published attack models and few researchers pay close attention to them. Just as figures shown, the RAdaBoost can not  effectively detect such attacks mainly because the present extractive features (as described in section 4)  are not enough to depict their material characteristics. Therefore,
the results  indicate the adaptive new classification features are needed for detecting such new attacks as PUA-AS, PUA-NR and PUA-ID.

\section{Conclusion and further discussions}

``Shilling'' attacks or ``profile injection'' attacks are serious threats to the collaborative filtering recommender systems (CFRSs).
 Since the number of detected attackers is far smaller than genuine users. Conventional supervised learning based detection methods have the challenges faced with this imbalanced classification. In the present paper, we improved the detection performance in two directions. First,  we extracted features from user profiles based on the statistical properties of the diverse attack models to make them much more easily classified. Then, we applied a variant of Boosting algorithm, called the re-scale AdaBoost (RAdaBoost) as our detection method, which
 gradually increasing emphasis on concerned attacks and could distinctly improve the predictive performance on a difficult classification task.
 And all our experimental results also demonstrated the outperformance of RAdaBoost in ``shilling'' attacks detection.

In our future work, we will explore more simpler and effective features to characterize attack profiles from different perspectives. The existing features based on basic description statistics and model-specific are difficult to fully discriminate between attack profiles and genuine profiles in diverse attack models. In addition, some features based on global calculating similarity such as DegSim (similarity with top neighbors) are unrealistic in mass user profiles, although they are effective to capture the concerned attack profiles. Therefore, how to extract local and effective features from user profiles is still an open issue.

%Besides,we will further validate the effectiveness of our proposed algorithm in other datasets, we will further design a series of experiments to address this task in our next work.
%Furthermore,
%To examine practically the effectiveness of our proposed method, a real-life application on product recommendation for large e-commerce company is also our further work.

\begin{figure}[H]
%\addtolength{\tabcolsep}{-3pt}
\centering
\subfigure{\label{Fig.sub.a}\includegraphics[height=20cm,width=14cm]{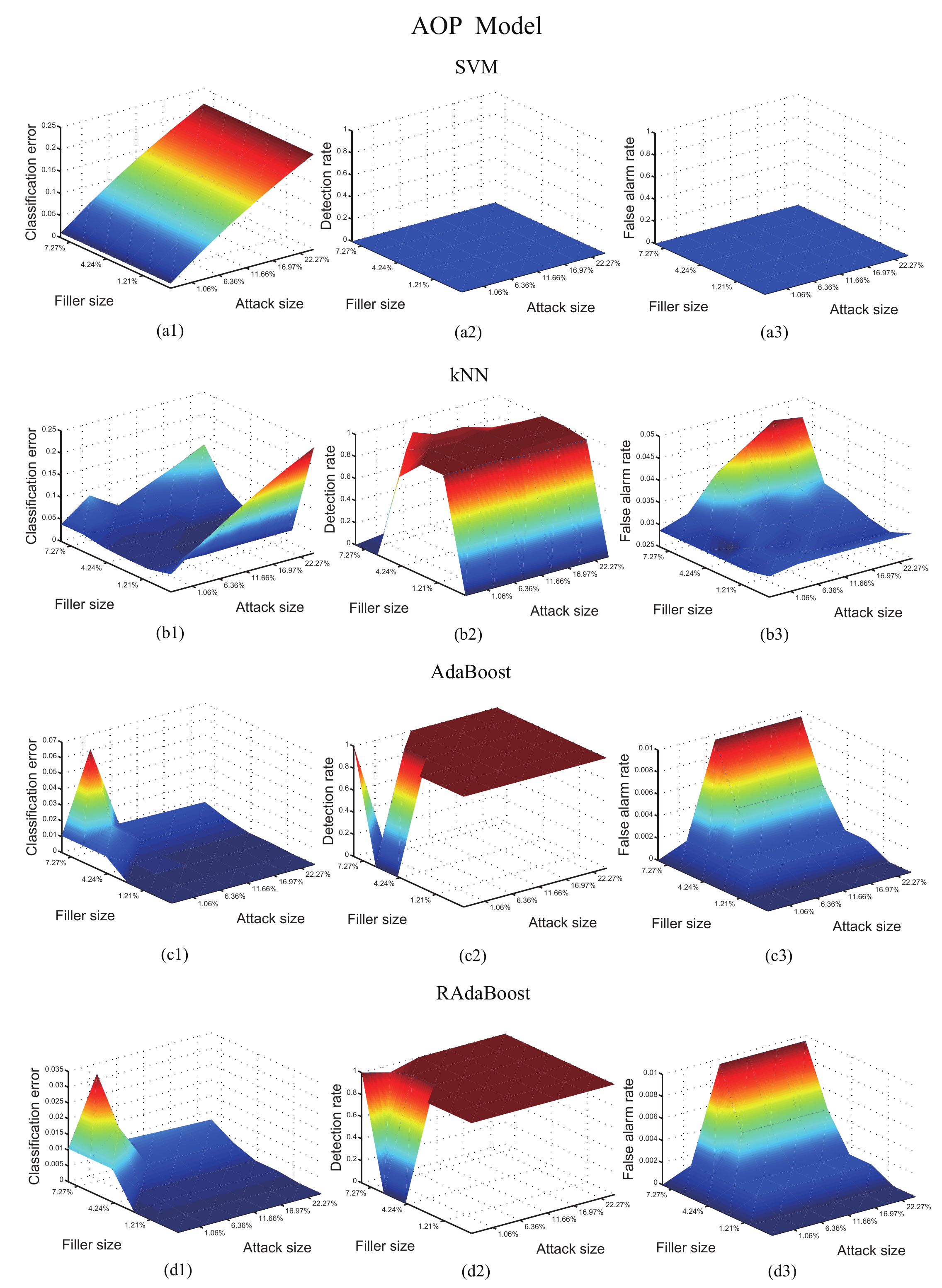}}
\caption{The classification error, detection rate and false alarm rate of RAdaBoost on different test sets in comparison with
  SVM, kNN and AdaBoost. AOP attack with diverse attack sizes \{1.1\%, 6.4\%, 11.7\%, 17.0\%, 22.3\%, 27.6\%\} and filler sizes \{1.2\%, 4.2\%, 7.3\%, 10.3\%, 13.3\%, 16.4\%\}.}
\label{f1}
\end{figure}
%\noindent

\begin{figure}[H]
%\addtolength{\tabcolsep}{-3pt}
\centering
\subfigure{\label{Fig.sub.a}\includegraphics[height=20cm,width=14cm]{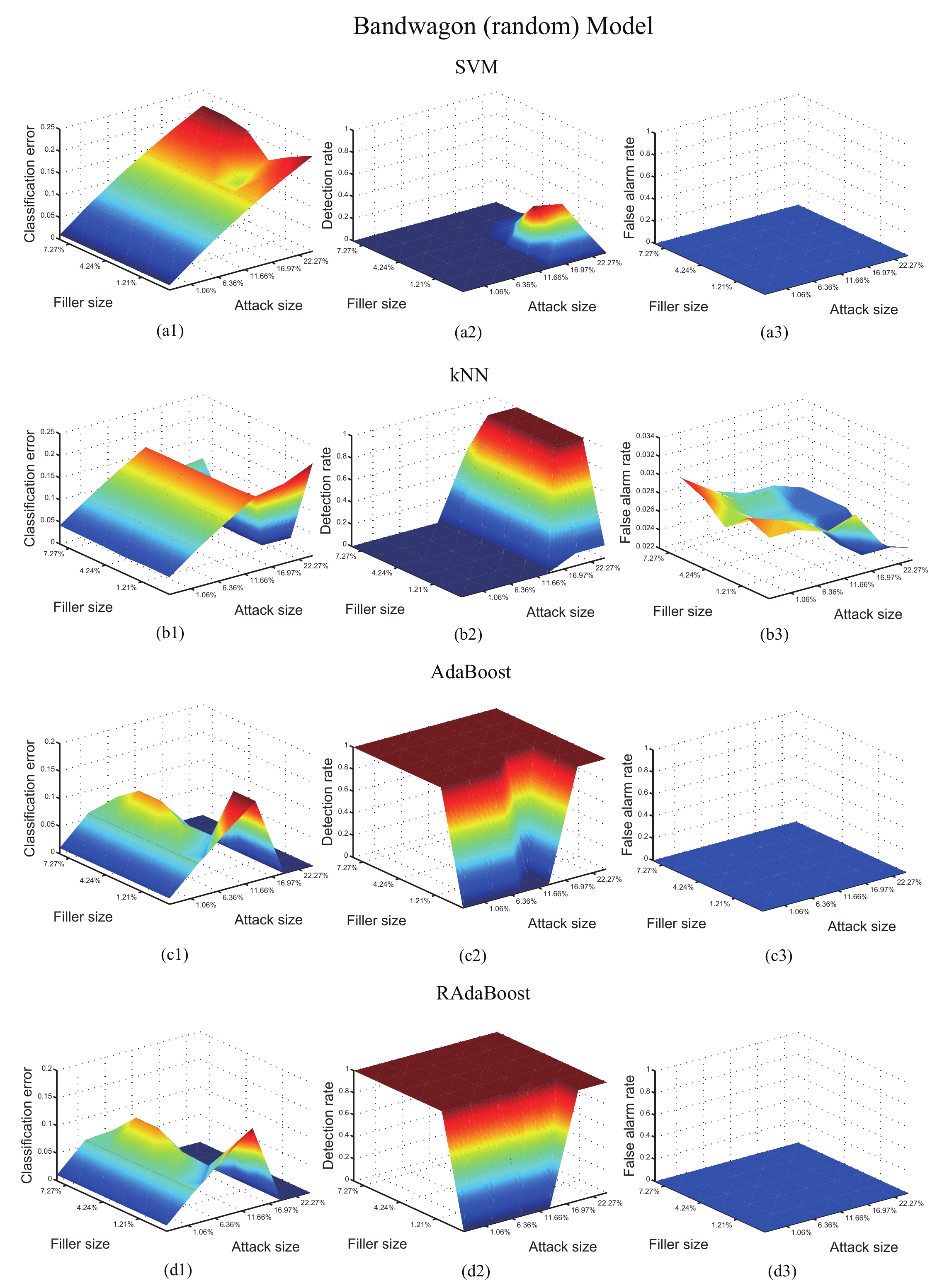}}
\caption{The classification error, detection rate and false alarm rate of RAdaBoost on different test sets in comparison with
  SVM, kNN and AdaBoost. Bandwagon (random) attack with diverse attack sizes \{1.1\%, 6.4\%, 11.7\%, 17.0\%, 22.3\%, 27.6\%\} and filler sizes \{1.2\%, 4.2\%, 7.3\%, 10.3\%, 13.3\%, 16.4\%\}.}
\label{f1}
\end{figure}

\begin{figure}[H]
%\addtolength{\tabcolsep}{-3pt}
\centering
\subfigure{\label{Fig.sub.a}\includegraphics[height=20cm,width=14cm]{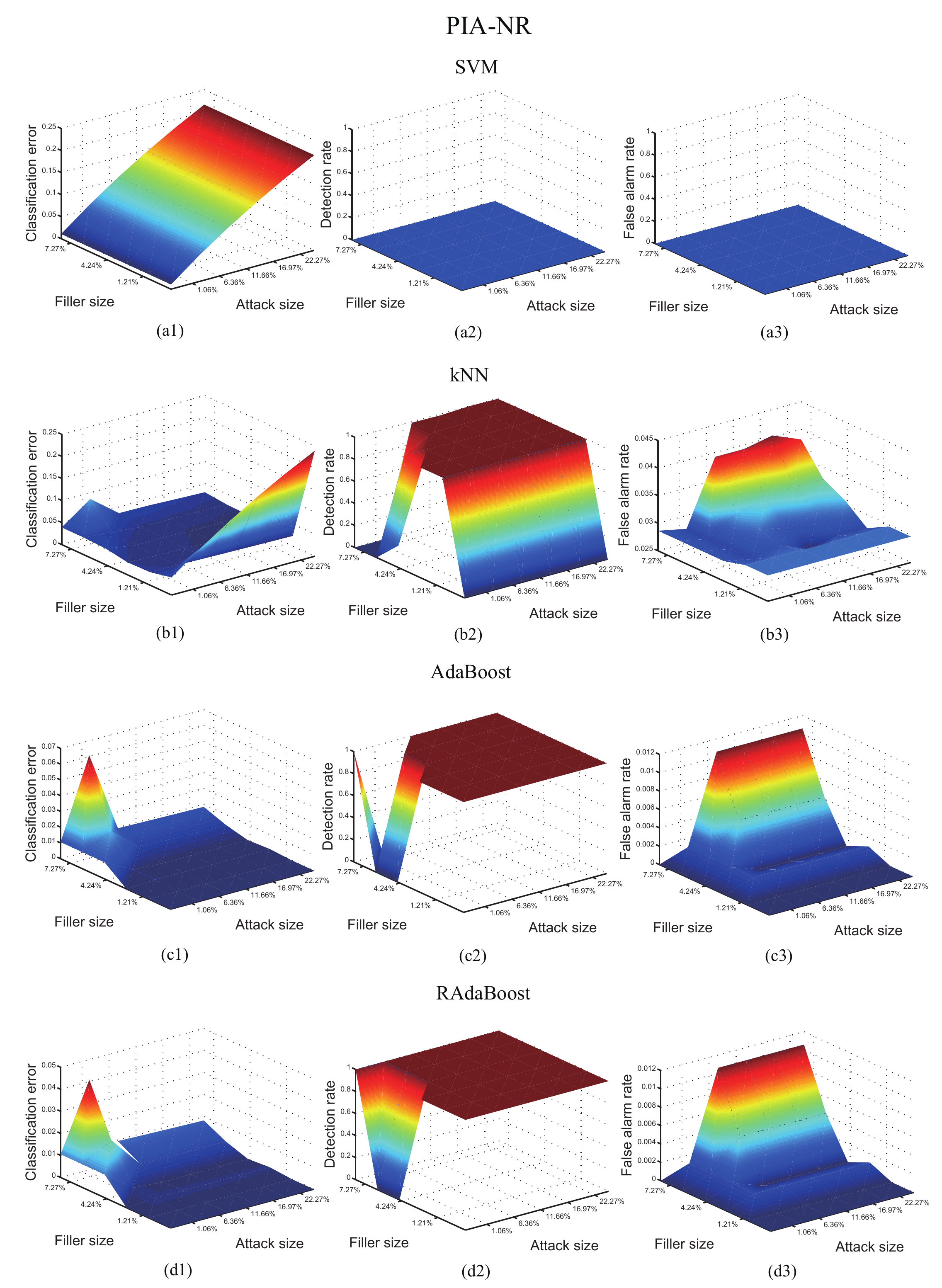}}
\caption{The classification error, detection rate and false alarm rate of RAdaBoost on different test sets in comparison with
  SVM, kNN and AdaBoost. PIA-NR attack with diverse attack sizes \{1.1\%, 6.4\%, 11.7\%, 17.0\%, 22.3\%, 27.6\%\} and filler sizes \{1.2\%, 4.2\%, 7.3\%, 10.3\%, 13.3\%, 16.4\%\}.}
\label{f1}
\end{figure}

\begin{figure}[H]
%\addtolength{\tabcolsep}{-3pt}
\centering
\subfigure{\label{Fig.sub.a}\includegraphics[height=20cm,width=14cm]{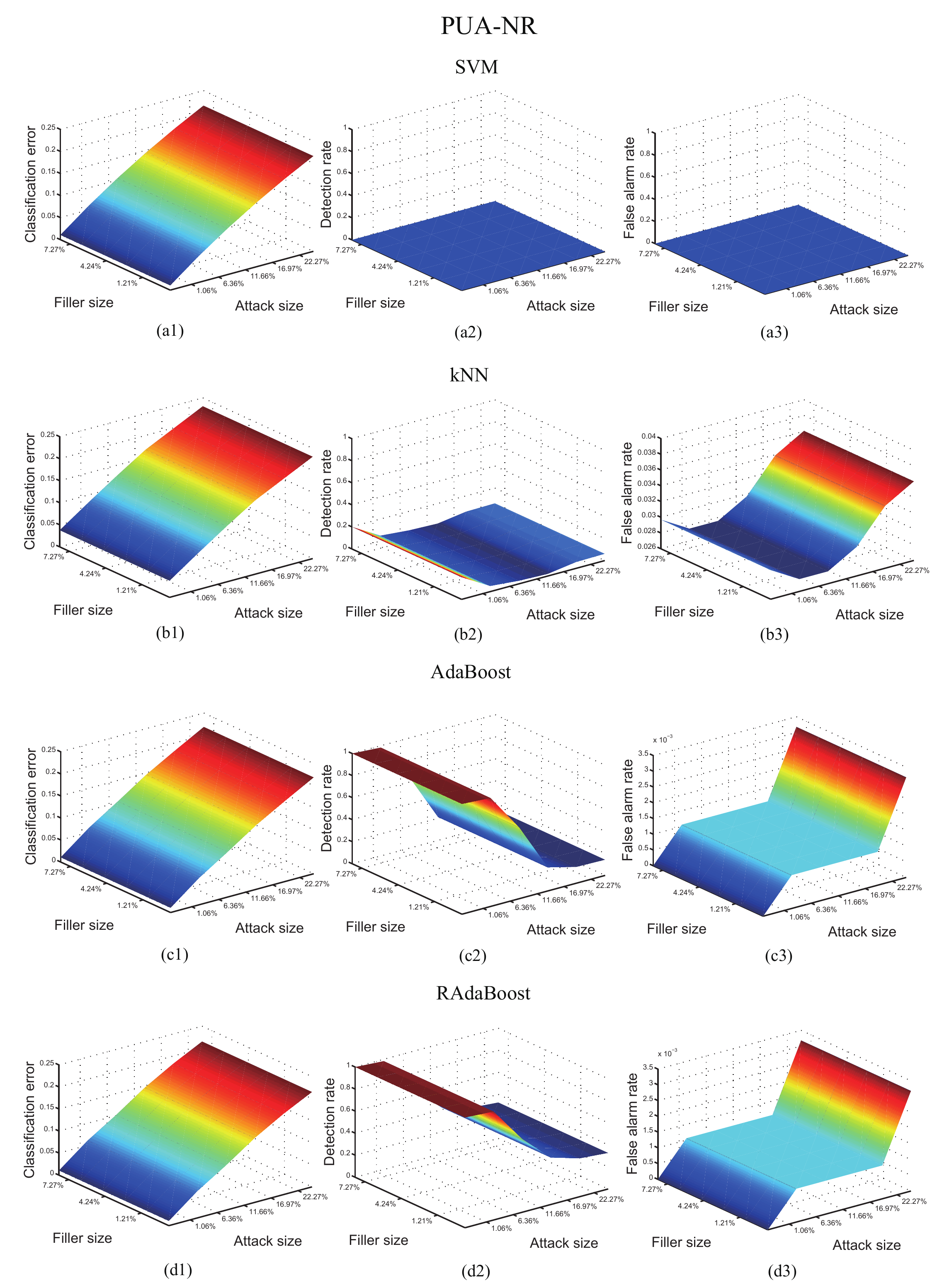}}
\caption{The classification error, detection rate and false alarm rate of RAdaBoost on different test sets in comparison with
  SVM, kNN and AdaBoost. PUA-NR attack with diverse attack sizes \{1.1\%, 6.4\%, 11.7\%, 17.0\%, 22.3\%, 27.6\%\} and filler sizes \{1.2\%, 4.2\%, 7.3\%, 10.3\%, 13.3\%, 16.4\%\}.}
\label{f1}
\end{figure}

%%For comparison, we can observe that the RBoosting can effectively detect all of the attacks except for PUA-AS and PUA-ID attacks, although it shows low detection rate when filler size is small. Notice that, RBoosting can fully detect ReverseBandwagon attack. In addition, to make comprehensive analysis on the PUA (PUA-AS, PUA-ID, and PUA-NR attacks), the performance surfaces of RBoosting shows poor detection rate with the attack size increasing. These results may indicate that fewer useful features of PUA attack profiles are extracted by the feature extraction methods.

\begin{figure}[H]
%\addtolength{\tabcolsep}{-3pt}
\centering
\subfigure{\label{Fig.sub.a}\includegraphics[height=20cm,width=14cm]{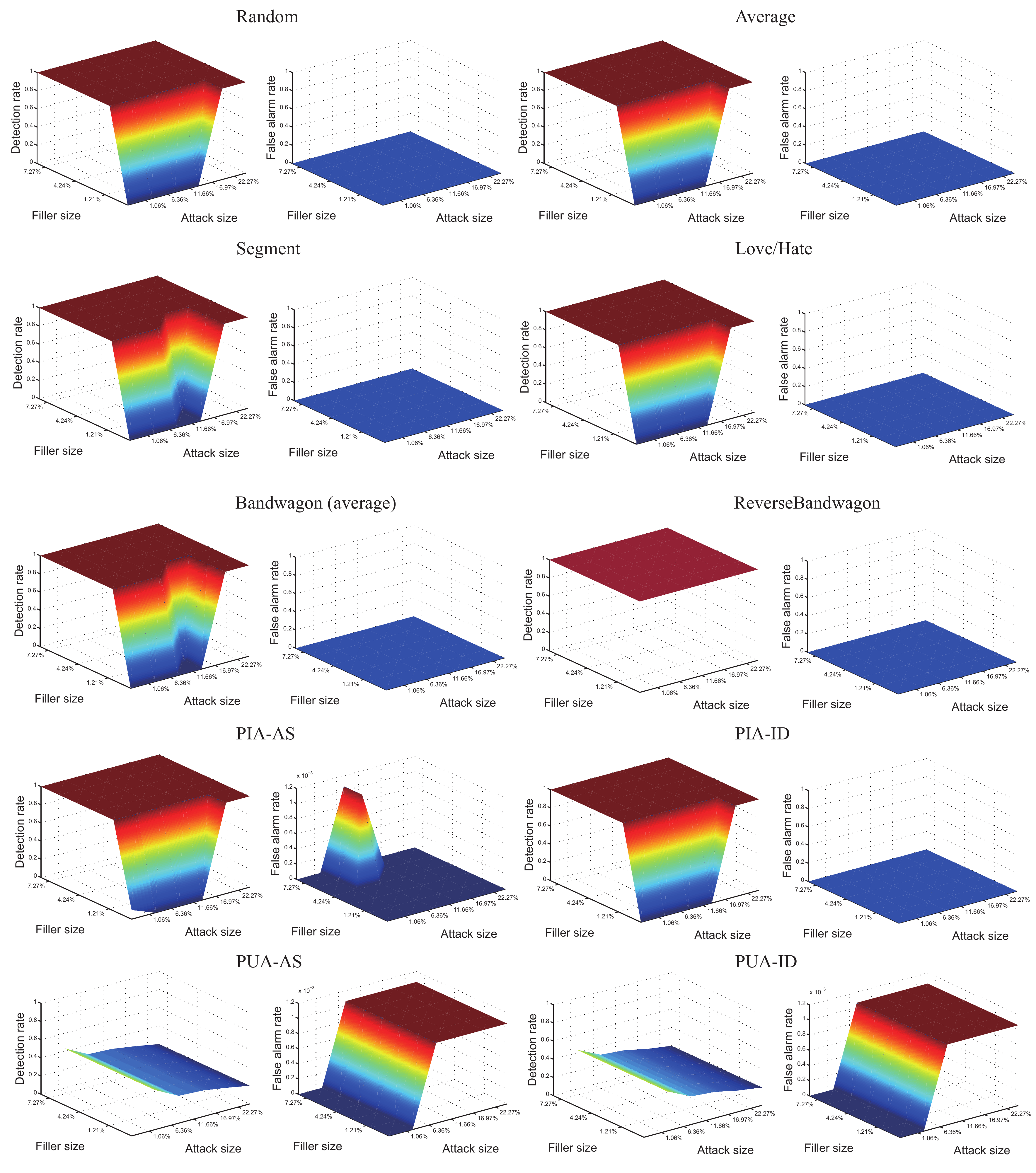}}
\caption{The detection rate and false alarm rate of RAdaBoost in 10 different attack models with diverse attack sizes \{1.1\%, 6.4\%, 11.7\%, 17.0\%, 22.3\%, 27.6\%\} and filler sizes \{1.2\%, 4.2\%, 7.3\%, 10.3\%, 13.3\%, 16.4\%\}.}
\label{f1}
\end{figure}

\bibliographystyle{plain}
\bibliography{KylinRef}

\end{document}